# The optimal design for cylindrical tubes used as acoustic stethoscopes for auscultation in COVID-19 diagnosis

Chuanyang Jiang[1], Jiaqi Zhao[2], Jiao Yu[3*]

[1]College of Mechanical Engineering, Liaoning Shihua University, Fushun, Liaoning Province, 113001, P. R. China.

[2]Department of Ultrasound, Changzheng Hospital, Second Military Medical University, Shanghai, 200003, P. R. China

[3]College of Science, Liaoning Shihua University, Fushun, Liaoning Province, 113001, P. R. China

E-mail: yujiaojoy@hotmail.com

**ABSTRACT**

During the COVID-19 outbreak, the auscultation of heart and lung sounds has played an important role in the comprehensive diagnosis and real-time monitoring of confirmed cases. With clinicians wearing protective clothing in isolation wards, a potato chip tube stethoscope, which is a secure and flexible substitute for a conventional stethoscope, has been used in the first-line treatment of COVID-19 by Chinese medical workers. In this study, an optimal design for this simple cylindrical stethoscope is proposed based on the fundamental theory of acoustic waveguides. Analyses of the cut-off frequency, sound power transmission coefficient, and sound wave propagation in the uniform lossless tube provide theoretical guidance for selecting the geometric parameters for this simple cylindrical stethoscope. In addition, relevant suggestions about surface treatments for the inner wall as well as the use of noise-reduction earplugs are also part of this optimal design.

**KEYWORDS:** COVID-19 pandemic; auscultation; potato-chips-tube stethoscope; optimal design; acoustic waveguide

# 1. INTRODUCTION

The COVID-19 pandemic has become an enormous public health concern, attracting intense attention not only in China but also around the globe. Given the high contagiousness of SARS-Coronavirus-2 (SARS-CoV-2),[1] strict personal protective measures have been implemented by the Chinese medical workers who have worked in infection isolation wards. In addition, the rigorous use of personal protective equipment, especially protective clothing, makes auscultation with conventional acoustic stethoscopes impossible.[2] However, it is essential to use stethoscopes to assess the subtleties of the heart and lung sounds and to monitor the progression of pneumonia with COVID-19 dynamically.

Under the circumstances, Gao[2] reported that a simple, disinfected cylindrical stethoscope made from an empty potato chip tube was applied during the first-line treatment of COVID-19. Though economical and safe, it is absolutely essential that this type of stethoscope, which is directly used in auscultation, is subjected to acoustic analysis and designed to obtain a more reliable acoustic performance. Hence, it is the aim of the present investigation to provide a theoretically optimal design for this alternative stethoscope.

# 2. ANALYTICAL STUDY

## A. Cut-off Frequency Analysis

In this study, a potato chip tube (Fig. 1) is modeled as a cylindrical rigid-walled waveguide with a diameter d = 6.530 cm. According to the basic theory of acoustic waveguides in constant cross section[3], the cut-off frequency of the plane wave in the hollow cylindrical waveguide is

$$f_c = \frac{1.841 c_0}{2\pi a} = \frac{1.841 \times 344}{\pi \times 6.530 \times 10^{-2}} = 3087.1 \, Hz, \tag{1}$$

where $a$ is the radius of the circular cross section and $c_0$ is the speed of sound in air at 20 °C. Considering that all the physiologically crucial heart and lung sounds range from 20 Hz to 1000 Hz,[4,5] which is evidently less than the $f_c$ in Eq. 1, the sounds used for auscultation will propagate only in plane-wave mode. Therefore, this pure propagation mode can greatly assist physicians and particularly cardiologists in dynamically monitoring as well as making accurate judgments regarding individual cardiopulmonary function.

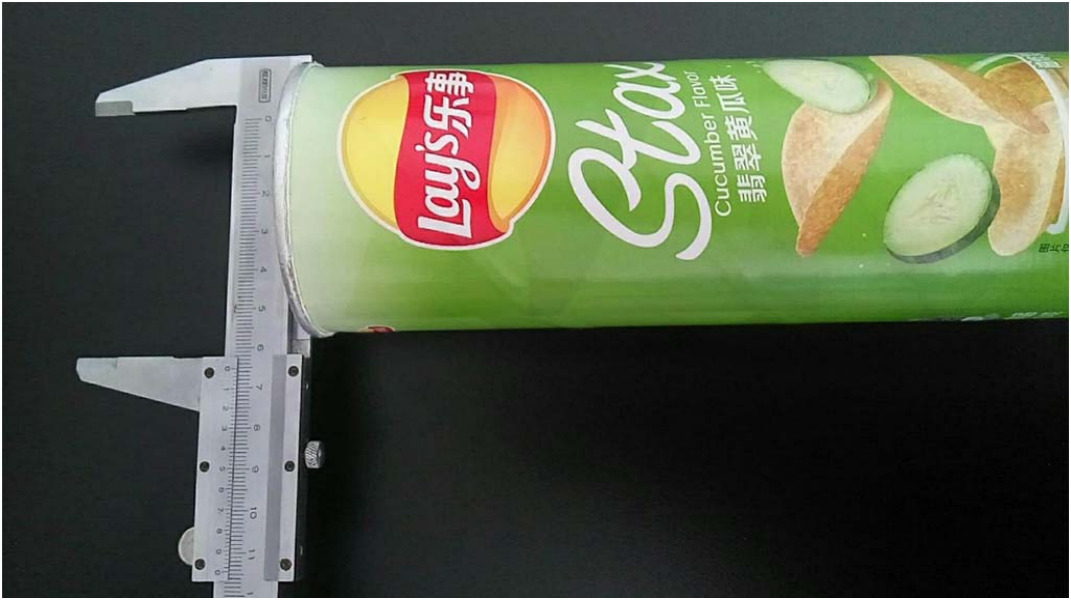

**Fig. 1.** A potato chip tube modeled as an acoustic waveguide when used as a simple stethoscope.

Notably, because the sounds that are valuable in the effective diagnosis of cardiac and pulmonary diseases by thoracic auscultation range from 20 Hz to 1000 Hz,[4,5] we were inspired to design the diameter dimension of the cylindrical stethoscope, and using Eq. 1, we obtain

$$\text{d} \leq \frac{1.841 c_0}{1000\pi} = 0.2\ m. \tag{2}$$

**B. Diameter Selection Analysis of the Cylindrical Stethoscope**

In addition to meeting the cut-off frequency requirements, the sound transmission to the external auditory canal must be considered in the design. To explore the essence of this research, slight differences in external auditory canal shapes and diameters between individuals are ignored. Moreover, because we are discussing the low frequency range below 1000 Hz, standing wave patterns in the human auditory canal do not need to be considered here because they occur at much higher frequencies[6-8].

Within the scope of the above approximation, the acoustic model that describes a circular waveguide with a sudden change in the cross-section is illustrated in Fig. 2 to analyze the sound transmission during auscultation.

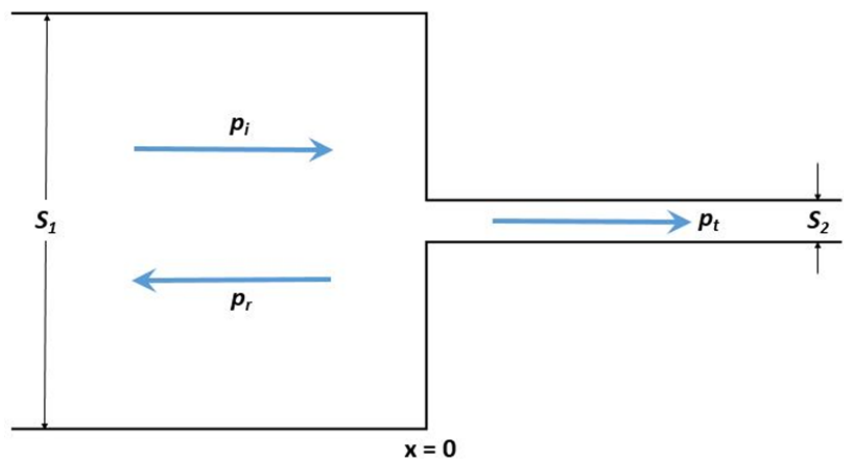

**Fig. 2.** Transmission and reflection of planar sound waves in the vicinity of a junction between two waveguides where the cross-sectional area changes from $S_1$ to $S_2$.

It is assumed that the coordinate origin is precisely at the junction between these two waveguides that represent the simple cylindrical stethoscope of cross-sectional area $S_1$ and a physician's external auditory canal of cross-sectional area $S_2$. It should be noted that the boundary conditions must include continuity in pressure and continuity in volume velocity. As a result, the sound power transmission coefficient in the simplified acoustic duct of the varied cross section[9] is

$$t_w = \frac{4\left(\frac{S_1}{S_2}\right)}{\left(\frac{S_1}{S_2}+1\right)^2}. \tag{3}$$

Using the first-order derivative of $t_w$ with regards to $\left(\frac{S_1}{S_2}\right)$ yields

$$\frac{d(t_w)}{d\left(\frac{S_1}{S_2}\right)} = \frac{4\left[1-\frac{S_1}{S_2}\right]}{\left(\frac{S_1}{S_2}+1\right)^3}. \tag{4}$$

From Eq. 4, the extreme point of function $t_w$ occurs when

$$\frac{S_1}{S_2} = 1, \tag{5}$$

and the second-order derivative of $t_w$ with regard to $\left(\frac{S_1}{S_2}\right)$ is negative, indicating that the sound power transmission coefficient reaches the largest value when $S_1$ equals $S_2$ in this application. In a real clinical situation, the size of the chest piece of the acoustic stethoscope, regardless of the diaphragms or the bells, is not as small as that of the cross-section of the external auditory canal (typically approximately 0.8 cm in diameter[10]), but it ranges from approximately 3.3 cm - 4.4 cm[11] due to a couple of factors that influence the picking up of sounds and vibration, including the effective auscultation area of heart and lung sounds as well as the curvature of the human body surface.

## C. Length Selection Analysis of the Cylindrical Stethoscope

For the optimal length of the cylindrical stethoscope, the fundamental equations that describe the sound wave propagation in the uniform, lossless waveguide are as follows[12]

$$-\frac{\partial p}{\partial x} = \frac{\rho}{A} \cdot \frac{\partial u}{\partial t}, \tag{6}$$

$$-\frac{\partial u}{\partial x} = \frac{A}{\rho c^2} \cdot \frac{\partial p}{\partial t}, \tag{7}$$

where $p$ is the sound pressure, $\rho$ is the density of air in the waveguide, A is the cross-sectional area, $c$ is the speed of sound in air, and u is the volume velocity at (x,t).

After using the Laplace transform, the transfer function of the cylindrical acoustic waveguide[12] is

$$V = \frac{2e^{-Sl/c}}{1+e^{-S \cdot \frac{2l}{c}}}, \tag{8}$$

where $= j\Omega$ ($\Omega$ is angular frequency). To calculate the formant frequency, when the denominator is zero,

$$1 + e^{-S \cdot \frac{2l}{c}} = 0, \tag{9}$$

we find

$$S = \pm \frac{j(2n+1)\pi c}{2l}. \quad (n=0, 1, 2, \ldots) \tag{10}$$

Then rejecting the negative value yields,

$$f = \frac{\Omega}{2\pi} = \frac{(2n+1)c}{4l}. \quad (n=0, 1, 2, \ldots) \tag{11}$$

First, we used $n = 0$ to discuss the fundamental frequency of the formant. From Ref. [2], the length of the commercially available potato chip tube ranges from 20.0 cm - 30.0 cm. When $l = 30$ cm, the fundamental frequency is

$$f_0 = \frac{344}{4 \times 0.3} \text{ Hz} = 286.7 \text{ Hz}. \tag{12}$$

Then, when $n = 1$ and $n = 2$, the values of the corresponding frequencies are $f_1 = 860$ Hz and $f_2 = 1433.3$ Hz. According to Eq. 12, to obtain a formant at a lower frequency, such as 100 Hz, a longer tube should be used.

## 3. DISCUSSION

Based on the analysis and calculation above, we can summarize the basic results on the optimal design of this simple cylindrical stethoscope. First, a diameter of less than 0.2 m will ensure the sound below 1000 Hz, which is physically important for cardiopulmonary function auscultation, to transmit in plane-wave mode within the cylindrical tube structure. Second, the diameter of this hollow cylindrical stethoscope is from 3.3 cm - 4.4 cm, which is believed to have a relative potential advantage over the current diameter of the potato chip tube stethoscope, partly because this size is suitable for picking up heart and lung sounds and partly because this design makes the difference between $S_1$ and $S_2$ smaller, contributing to the sound transmission to the external auditory canal. Third, the use of the current length (30 cm) of the cylindrical stethoscope should cover the medium and high frequency ranges of physically important heart and lung sounds (300 Hz - 1000 Hz). In using Eq. 11 at n = 0, we derive that if the length of the cylindrical stethoscope increases to 0.8 m - 0.9 m, the low frequency heart sounds of 95.6 Hz - 107.5 Hz will be amplified via the formant effect, to reach a level that is easy to hear. However, given the flexibility and convenience of the device operation, lengths longer than 0.9 m are not recommended during real auscultation when using this simple diagnostic tool.

Moreover, instead of using the original material from the inner wall of the cylindrical stethoscope, which is to some degree infiltrated by oil, an appropriate substitute, such as a coat of enamel, should be painted smoothly and uniformly on the surface to reduce the frictional loss of sound due to viscosity. In addition, a great deal of attention must be paid to the fact that medical workers who intend to conduct auscultations using this simple device ought to wear a single earplug in the ear that is not being used during auscultation to reduce the impact of ambient noise during diagnosis.

Dating back to 1816, the French doctor Laennec, who invented the world's first stethoscope, successfully conducted a cardiac auscultation using a cylindrical paper stethoscope.[13,14] After continuous optimization and improvement, modern stethoscopes now possess better acoustic performance during cardiac and pulmonary auscultation. Nevertheless, strict personal protective equipment use makes auscultation using modern stethoscopes impossible in diagnosing confirmed COVID-19 cases. Given the pressing situation, this simple cylindrical stethoscope, which has advantages similar to those of Laennec's original invention, inspires us to complete the acoustic calculation and design herein and to reflect on the reductionism as well as pragmatism in the design and use of diagnostic tools during the COVID-19 pandemic.

Last but not least, it should be noted that the above calculation and analysis of the fundamental frequency and the high-order formant frequency are based on the basic theory of acoustic waveguides. Therefore, further experimental investigations should be performed by practitioners in clinical settings to validate the sensitivity and applicability of our improved simple stethoscope.

## 4. CONCLUSIONS

In conclusion, during the COVID-19 pandemic, simple cylindrical stethoscopes have been applied to cardiac and pulmonary function auscultation in China, and this economical instrument is considered as a suitable diagnostic tool when conventional stethoscopes cannot be used by medical workers in protective clothing, or when medical resources are extremely scarce in poverty-stricken areas of developing countries. The cut-off frequency analysis indicates that a hollow cylindrical tube with a diameter of less than 0.2 m allows the heart and lung sounds within the auscultation frequency range to propagate in plane-wave mode. Compared with the potato chip tube stethoscope, we propose an optimization of this simple stethoscope by reducing the diameter to 3.3 cm - 4.4 cm. For a simple cylindrical stethoscope measuring 30 cm in length, the medium and high frequencies of the heart and lung sounds can induce acoustic resonance when transmitted down the tubing of the simple stethoscope which is considered to be beneficial for auscultation. The acoustic performance of the simple stethoscope for low-frequency chest sounds requires further experimental confirmations from clinicians, and it could be improved by increasing its length.

## 5. ACKNOWLEDGMENTS

This work was financially supported by the National Natural Science Foundation of China (81501492), the Natural Science Foundation of Liaoning Province of China (2019-MS-219), the Natural Science Foundation of Shanghai of China (20ZR1457900), Liaoning Revitalization Talents Program (XLYC1907034).

**REFERENCES AND LINKS**

**Figure Captions**

**Fig. 1.** A potato chip tube modeled as an acoustic waveguide when used as a simple stethoscope

**Fig. 2.** Transmission and reflection of planar sound waves in the vicinity of a junction between two waveguides where the cross-sectional area changes from $S_1$ to $S_2$